\documentclass[11pt]{article} 

\usepackage{times} 
\usepackage{graphicx} 
\usepackage{amsmath}

\title{Brownian motion and gambling:\\ from ratchets to paradoxical games}
\author{J.M.R. Parrondo \and  Luis Din\'{\i}s \\
Grupo Interdisciplinar de Sistemas Complejos (GISC) and
\\Dept.~de F\'{\i}sica At\'{o}mica,
 Molecular y Nuclear,\\
 Universidad
Complutense de Madrid, 28040-Madrid, Spain.}

\begin{document}
\maketitle

\begin{abstract}
Two losing gambling games, when alternated in a periodic or random
fashion, can produce a winning game. This paradox has been
inspired by certain physical systems capable of rectifying
fluctuations: the so-called Brownian ratchets. In this paper we
review this paradox, from Brownian ratchets to the most recent
studies on collective games, providing some intuitive explanations
of the unexpected phenomena that we will find along the way.
\end{abstract}

\section{Introduction: A noisy revolution}

Imagine two simple gambling games, say A and B, in which I play
against you. Each one is a losing game for me, in the sense that
my average capital is a decreasing function of the number of turns
we play. Once you are convinced that I lose in both games, I give
you a third proposal: alternate the games following the sequence
AABBAABB... If you frown, the proposal can be modified to make it
less suspicious: in each run we will randomly chose the game that
is played. If you accept either of these proposals you would have
trusted your intuition too much, not realising that random systems
may behave in an unexpected way.

The phenomenon we have just described is known as {\em Parrondo's
paradox} \cite{abbott,abbott-nature,abbottrev}. It was originally
inspired by a class of physical systems: {\em the Brownian
ratchets} \cite{ajdari,magnasco,astumian,reiman,linke} and lately
has received the attention of scientists working on several
fields, ranging from biology to economics. These are systems
capable of rectifying thermal fluctuations, such as those
exhibited by a Brownian particle.

Brownian motion was one of the first crucial proofs of the
discreteness of matter. First observed by Jan Ingenhousz in 1785,
and later rediscovered by Brown in 1828, the phenomenon consists
on the erratic or fluctuating motion of a small particle when it
is embedded in a fluid. In the beginning of the XXth century,
Einstein realized that these fluctuations were a manifestation of
the molecular nature of the fluid\footnote{The thermal origin of
Brownian motion was firstly proposed by Delsaux in 1877 and later
on by Gouy in 1888 (see \cite{einstein}).} and devised a method to
measure Avogadro's number by using Brownian motion
\cite{einstein}. Since then, the study of fluctuations has been a
major topic in statistical mechanics.

The theory of fluctuations helped to understand noise in
electrical circuits, activation processes in chemistry, the
statistical nature of the second law of thermodynamics, and the
origin of critical phenomena and spontaneous symmetry breaking, to
cite only a few examples. In most of these cases, the role played
by thermal fluctuations or thermal noise is either to trigger some
process or to act as a disturbance. However, in the past two
decades, the study of fluctuations has led to models and phenomena
where the effect of noise is more complex and sometimes unexpected
and even counterintuitive.

Noise can enhance the response of a nonlinear system to an
external signal, a phenomenon known as {\em stochastic resonance}
\cite{hanggires}. It can create spatial patterns and ordered
states in spatially extended systems \cite{nipt,jordi}, and
Brownian ratchets show that noise can be rectified and used to
induce a systematic motion in a Brownian particle
\cite{ajdari,magnasco,astumian,reiman,linke}. In these new
phenomena, noise has a very different role from that considered in
the past: it contributes to the creation of order. This could be
relevant in several fields, and specially in biology, since most
biological systems manage to keep themselves in ordered states
even while
 surrounded by noise, both thermal noise at the level of the cell
and environmental fluctuations at the macroscopic level.

However, fluctuations are not only restricted to physics,
chemistry or biology. The origin of the theory of probability is
closely related to gambling games, social statistics, and even to
the efficiency of juries \cite{hacking}. Statistical mechanics and
probability theory have both contributed to each other and also to
fields like economics. In 1900, five years before Einstein's
theory of the Brownian motion, the French mathematician Louis
Bachelier worked out a theory for the price of a stock very
similar to Einstein's \cite{bachelier}. Recently this link between
probability, statistical mechanics, and economics has crystallised
in a new field: {\em econophysics} \cite{econophys}.

Some of the aforementioned constructive role of noise has been
observed in complex systems beyond physics. Stochastic resonance,
for instance, has an increasing relevance in the study of
perception and other cognitive processes \cite{hanggires,prlres}.
Similarly, we expect that other elementary stochastic phenomena
such as rectification will be observed in many situations not
restricted to physics.

With this idea in mind, Parrondo's paradox came up as a
translation to simple gambling games of a Brownian ratchet
discovered by Ajdari and Prost \cite{ajdari}. The ratchet was
afterwards named by Astumian and Bier the {\em flashing ratchet}
\cite{astumian} and it was related to the idea proposed by
Magnasco \cite{magnasco} that biological systems could rectify
fluctuations to perform work and systematic motion.

The paradox does not make use of Brownian particles, but only of
the simpler fluctuations arising in a gambling game. However,  it
illustrates the mechanism of rectification in a very sharp way,
and for this reason we think that it could contribute to extend
the ``noisy revolution", i.e., the idea that noise can create
order, to those fields where stochastic dynamics is relevant.

The paper is organised as follows. In section \ref{sec:ratchets}
we briefly review the flashing ratchet and explain how it can
rectify fluctuations. Section \ref{sec:games} is devoted to the
original Parrondo's paradox. There we introduce the paradoxical
games as a discretisation of the flashing ratchet, discuss an
intuitive explanation of the paradox that we have called {\em
reorganisation of trends}, and present an extension of the
original paradox inspired by this idea. In section
\ref{sec:colect} we introduce several versions of the games
involving a large number of players. Some interesting effects can
be observed in these collective games: redistribution of capital
brings wealth \cite{capital}, and collective decisions taken by
voting or by optimizing the returns in the next turn can lead to
worse performance than purely random choices
\cite{eplsr,physicaA}. Finally, in section \ref{sec:conclusions}
we briefly review the literature on the paradox and present our
main conclusions.

\section{Ratchets}
\label{sec:ratchets}

Here we  revisit the flashing ratchet \cite{ajdari,astumian}, one
of the simplest Brownian ratchets and the most closely related to
the paradoxical games. We refer to the exhaustive review by
Reimann on Brownian ratchets \cite{reiman} or the special issue in
Applied Physics A, edited by Linke \cite{linke}, for further
information on the subject.

\begin{figure}
\begin{center}
\includegraphics[height=0.3\textheight]{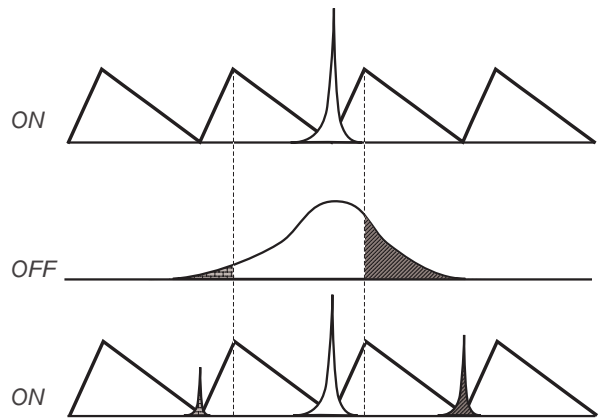}
\caption{ The flashing ratchet at work. The figure represent three
snapshots of the potential and the density of particles. Initially
(upper figure), the potential is on and all the particles are
located around one of the minima of the potential. Then the
potential is switched off and the particles diffuse freely, as
shown in the centered figure, which is a snapshot of the system
immediately before the potential is switched on again. Once the
potential is connected again, the particles in the darker region
move to the right hand minimum whereas those within the small grey
region move to the left. Due to the asymmetry of the potential,
the ensemble of particles move, in average, to the right.}
 \label{figflash}
\end{center}
\end{figure}

Consider an ensemble of independent one-dimensional Brownian
particles in the asymmetric sawtooth potential depicted in
Fig.~\ref{figflash}. It is not difficult to show that, if the
potential is switched on and off periodically, the particles
exhibit an average motion to the right. Let us assume that the
temperature $T$ is low enough to ensure that $kT$ is much smaller
than the maxima of the potential, and that we start with the
potential switched on and with all the particles around one of its
minima, as shown in the upper plot of Fig.~\ref{figflash}. When
the potential is switched off, the particles diffuse freely, and
the density of particles spreads as depicted in the central plot
of the figure. If the potential is then switched on again, each
particle will move back to the initial minimum or to one of the
nearest neighboring minima, depending on its position. Particles
within the dark region will move to the right hand minimum, those
within the small grey region will move to the left hand minimum,
and those within the white region will move back to their initial
positions. As is apparent from the figure and due to the asymmetry
of the potential,  more particles fall into the right hand
minimum, and thus there is a net motion of particles to the right.
For this to occur, the switching can be either random or periodic,
but the average period must be of the order of the time to reach
the nearest barrier by free diffusion (see \cite{ajdari,astumian}
for details).

This motion can be seen as a rectification of the thermal noise
associated with  free diffusion. The diffusion is symmetric: some
particles move to the right and some to the left, but their
average position does not change. However, when the potential is
switched on again, most of the particles that moved to the left
are driven back to the starting position, whereas many particles
that moved to the right are pushed to the right hand minimum. The
asymmetric potential acts as a rectifier: it ``kills" most of the
negative fluctuations and  ``promotes" most of the positive ones.

The effect remains if we add a small force toward the left, i.e.,
in a direction opposite to the induced motion. In this case,  the
ratchet still induces a motion against the force. Consequently,
particles perform work, and the system can be considered a
Brownian motor. It can be proved that this type of motor is
compatible with the second law of thermodynamics. In fact, the
efficiency of such a motor is far below the limits imposed by the
second law \cite{energrev,europhys}. However, the ratchet with a
force exhibits a curious property: when the potential is
permanently on or off, the Brownian particles move in the same
direction as the force, whereas they move in the opposite
direction when the potential is switched on and off. This is the
essence of the paradoxical games: we have two dynamics; in each
one a quantity, namely the position of the Brownian particle,
decreases in average; however, the same quantity increases in
average when the two dynamics are alternated.

\section{Games}
\label{sec:games}

The flashing ratchet can be discretised in time and space, keeping
most of its interesting features. The discretised version adopts
the form of a pair of simple gambling games, which are the basis
of the Parrondo's paradox.

\subsection{The original paradox}

We consider two games, A and B, in which a player can make a bet
of 1 euro. $X(t)$ denotes the capital of the player, where
$t=0,1,2\ldots$ stands for the number of turns played. Game A
consists of tossing a slightly biased coin so that the player has
a probability $p_A$ of winning which is less than a half. That is,
$p_A=1/2-\epsilon$, where the bias $\epsilon$ is a small positive
number.

\begin{figure}
\begin{center}
\includegraphics[height=0.28\textheight]{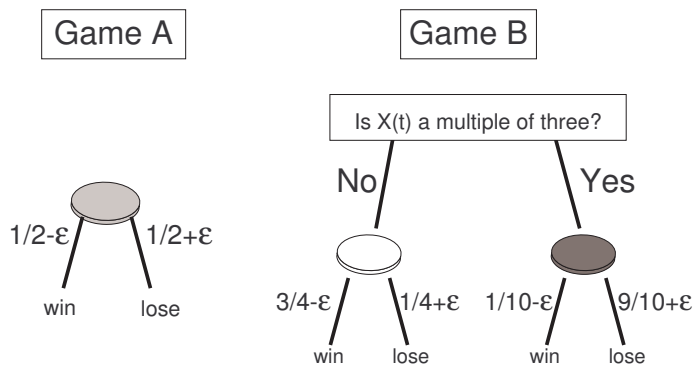}
\caption{Rules of the paradoxical games. In game A, the player
wins (her capital increases by one euro) with a probability
$1/2-\epsilon$ and loses (her capital decreases by one euro) with
a probability $1/2+\epsilon$, $\epsilon$ being a small positive
number. In the figure, these probabilities are represented by a
coin with two possible outcomes. In game B, the probability to win
and lose depends on the capital $X(t)$ of the player: if $X(t)$ is
a multiple of three, then we use a ``bad" coin, with a probability
to win equal to $1/10-\epsilon$; if $X(t)$ is not a multiple of
three, then a ``good" coin, with a probability to win equal  to
$3/4-\epsilon$, is used. In the figure the darkness of the coins
represents their ``badness'' for the player.}
 \label{figrules}
\end{center}
\end{figure}

The second game, B, is played with two biased coins, a ``bad
coin'' and a ``good coin''. The player must toss the bad coin if
her capital $X(t)$ is a multiple of $3$, the probability of
winning being $p_{\text{bad}}=1/10-\epsilon$. Otherwise, the good
coin  is tossed and the probability of winning is
 $p_{\text{good}}=3/4-\epsilon$. The rules of games A and B are
 represented in Fig.~\ref{figrules}, in which the darkness
 represents the ``badness" of each coin.

For these choices of $p_A,p_{\text{good}} \text{ and }
 p_{\text{bad}}$, both games are fair if $\epsilon=0$, in the sense
 that $\langle X(t)\rangle$ is constant. This is evident for game
 A, since the probabilities to win and lose are equal. The analysis of game B
 is more involved, but we will soon prove that the effect of the good
 and the bad coins cancel each other for $\epsilon=0$.

\begin{figure}
\begin{center}
\includegraphics[width=0.65\textwidth,]{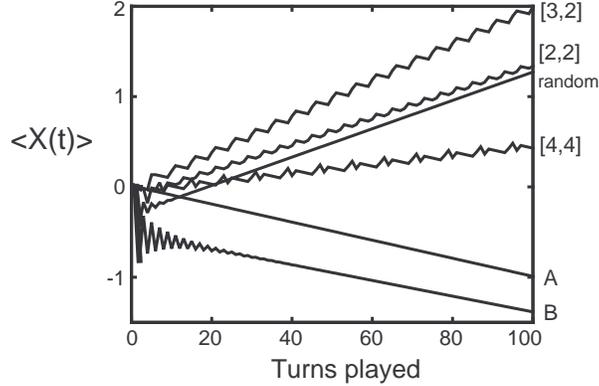}
\caption{Average capital for 5000 players as a function of the
number of turns for game A, B and their periodic and random
combinations. $\epsilon=0.005$ and $[a,b]$ stands for periodic
sequences where A (B) is played $a$ ($b$) consecutive turns.}
\label{paradoja}
\end{center} \end{figure}

On the other hand, both games have a tendency to lose if
$\epsilon>0$, i.e.,  $\langle X(t)\rangle$ decreases with the
number of turns $t$. Surprisingly enough, if the player randomly
chooses the game to play in each turn, or plays them following
some predefined periodic sequence such as ABBABB..., then her
average capital $\langle X(t) \rangle$ is an increasing function
of $t$, as can be seen in Fig.~\ref{paradoja}.

\begin{figure}
\begin{center}
\includegraphics[width=0.55\textwidth]{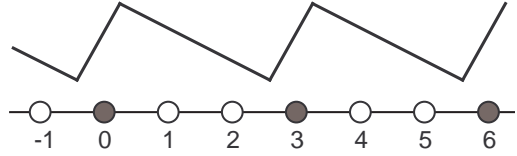}
\caption{ A random walk picture of game B compared with the
ratchet potential. The bad coin (black dots) plays the role of the
negative force acting on a short interval, whereas the two
consecutive good coins (white dots) are the analogous of the
positive force acting on the long intervals.} \label{figrw}
\end{center} \end{figure}

The paradox is closely related to the flashing ratchet. If we
visualise the capital $X(t)$ as the position of a Brownian
particle in a one dimensional lattice, game A, for $\epsilon=0$,
is a discretisation of the free diffusion, whereas game B
resembles the motion of the particle under the action of the
asymmetric sawtooth potential. Fig. \ref{figrw} shows this spatial
representation for game B compared with the ratchet potential.
When the particle is on a dark site, the bad coin is used and the
probability to win is very low, whereas on the white sites the
most likely move is to the right. The sawtooth potential has a
short spatial interval in which the force is negative and a long
interval with a positive force. Equivalently, game B uses a bad
coin on a ``short interval'', i.e., on one site of every three on
the lattice, and a good coin on a ``long interval'' corresponding
to two consecutive sites which are not multiple of three (see Fig.
\ref{figrw}).

As in the flashing ratchet, game B rectifies the fluctuations of
game A. Suppose that we play the sequence AABBAABB... and that
$X(t)$ is a multiple of three immediately after two instances of
game B. Then we play game A twice, which can drive the capital
back to $X(t)$ or to $X(t)\pm 2$. In the latter case, the next
turn is for game B with a capital that is not a multiple of three,
which means a good chance of winning. That is, game B rectifies
the fluctuations that occurred in the two turns of game A. The
rectification is not as neat as in the low temperature flashing
ratchet, but enough to cause the paradox.

There is a more rigorous way of associating a potential to a
gambling game by using a master equation \cite{raulfp}. However,
it provides a similar picture of game B, as a random walk that is
nonsymmetric under inversion of the spatial coordinate and capable
of rectifying fluctuations.

\subsection{Reorganisation of trends}
\label{closerlook}

Beside the ratchet effect, one can explain the paradox considering
another interesting mechanism. Recall that game B is played with
two coins: a good one, used whenever the capital of the player is
not a multiple of three, and a bad one which is used when the
capital is a multiple of three. Therefore, the ``profitability" of
game B crucially depends on how often the bad coin is used, i.e.,
on the probability $\pi_0$ that the capital is a multiple of
three. It turns out that, when game B is played, this probability
is not $1/3$, as one could naively expect, but larger. This can be
reasoned from figure \ref{figrw}. When the capital is at a white
site, its most likely move is to the right, whereas at dark sites
the most likely move is to the left. The capital thus spends more
time jumping forth and back between a multiple of three and its
left hand neighbour than what would do if it moved completely at
random. Consequently, the probability $\pi_0$ is larger than 1/3.
On the other hand, under game A the capital does move in a random
way. Therefore, playing game A in some turns shifts $\pi_0$
towards 1/3, or, equivalently, reduces the number of times the bad
coin of game B is used. In other words, game A, although losing,
boosts the effect of the good coin in B, giving the overall game a
winning tendency. We have named this mechanism {\em reorganization
of trends}, since game A reinforces the positive trend already
present in game B.

Along this section, we  formulate this argument in a quantitative
way. Let us first consider game B separately. The probability to
win in the $t$-th turn can be calculated as
\begin{equation}
p_{\text{win}}(t)=\pi_0(t)p_{\text{bad}}+\left[
1-\pi_0(t)\right]p_{\text{good} }\label{pganar}
\end{equation}
where $\pi_0(t)$ is the probability of $X(t)$ being a multiple of
3 (i.e. of using the bad coin).

One can calculate the value of $\pi_0(t)$, by using very simple
techniques from the theory of Markov chains. First, we define the
random process
\begin{equation} Y(t)\equiv X(t)\mod 3
\end{equation}
taking on only three possible values or states, $Y(t)=0,1,2$,
depending on whether the capital $X(t)$ is a multiple of three, a
multiple of three plus one, or a multiple of three plus two,
respectively. This variable $Y(t)$ is a Markov process, i.e., the
statistical properties of $Y(t+1)$ depend only on the value taken
on by $Y(t)$. This allows one to derive a {\em master equation}
for its probability distribution.

Let $\pi_0(t),\pi_1(t),\pi_2(t)$ be the probability that $Y(t)$ is
equal to 0, 1, and 2, respectively. There are two possibilities
for $Y(t)=2$ to occur: either $Y(t-1)=0$ and we lose in the $t$-th
turn (with probability $1-p_{\rm bad}$), or $Y(t-1)=1$ and we win
in the $t$-th turn (with probability $p_{\rm good}$). Therefore:
\begin{equation} \pi_2(t)=(1-p_{\rm bad})\pi_0(t-1)+p_{\rm
good}\pi_1(t-1).
\end{equation}

Following the same type of argument, one can derive equations for
$\pi_0(t)$ and $\pi_1(t)$, and the three equations can be written
in matrix form as: \begin{equation} \label{equation13}
\vec{\pi}(t)=\Pi_B\vec{\pi}(t-1)
\end{equation}
where
\begin{equation}\vec{\pi}(t)\equiv  \left(
\begin{array}{c} \pi_0(t) \\ \pi_1(t) \\
\pi_2(t)\end{array}\right)
\end{equation}
and
\begin{equation}
\Pi_B\equiv \left(
\begin{array}{ccc} 0 & 1-p_{\rm good} & p_{\rm good}\\
p_{\rm bad} & 0 & 1-p_{\rm good} \\ 1-p_{\rm bad} & p_{\rm good} &
0
\end{array} \right)\label{matrizB}.
\end{equation}

After a small number of turns of game B, $\vec\pi(t)$ approaches
to a stationary value $\vec\pi^{\rm st}_{\rm B}$, which is
invariant under the transformation given by Eq.
(\ref{equation13}), i.e.:
\begin{equation}
\vec\pi^{\rm st}_{\rm B}=\Pi_B\vec\pi^{\rm st}_{\rm B}.
\end{equation}
The third component of the solution of this equation reads:
 \begin{equation}
\pi_{\rm 0B}^{\rm
st}=\frac{5}{13}-\frac{440}{2197}\epsilon+O(\epsilon^2 )\simeq
0.38 - 0.20\,\epsilon \label{piob}
\end{equation}
where we have used the values of the original paradox, $p_{\rm
bad}=1/10-\epsilon$ and $p_{\rm good}=3/4-\epsilon$, and have
expanded the solution up to first order of $\epsilon$, to simplify
the exposition.

Substituting this value in Eq.~(\ref{pganar}) we obtain the
probability of winning for game B for sufficiently large $t$
\begin{equation}
p_{\text{win,B}}=\frac{1}{2}-\frac{147}{169}\epsilon+O(\epsilon^2)
\end{equation}
which is less than $1/2$ for $\epsilon>0$. This proves that game B
is fair for $\epsilon=0$ and losing for $\epsilon>0$, as shown in
Fig.~\ref{paradoja}.

The paradox arises when game A comes into play. Game A is always
played with the same coin, regardless of the value of the capital
$X(t)$, and therefore drives the probability distribution
$\vec\pi(t)$ to a uniform distribution. Thus, game A makes
$\pi_0(t)$ tend to $1/3$. Since $1/3< \pi_{\rm 0B}^{\rm st}$, the
effect of game A is to decrease the probability of using the bad
coin in the turns where B is played.

This can be seen in a more precise way, since the random
combination of games A and B can be again solved by using the
master equation:
\begin{equation}
\vec\pi^{\rm st}_{\rm
AB}=\frac{1}{2}\left[\Pi_B+\Pi_A\right]\vec\pi^{\rm st}_{\rm AB}
\end{equation}
where
\begin{equation}
\label{mat-a} \Pi_A= \left(\begin{array}{ccc}
  0 & 1-p_A & p_A \\
  p_A & 0 & 1-p_A \\
  1-p_A & p_A & 0 \\
\end{array}\right)
\end{equation}
with $p_A=1/2-\epsilon$. The probability of using the bad coin
decreases to
\begin{equation}
\pi^{\rm st}_{\rm
0AB}=\frac{245}{709}-\frac{48880}{502681}\epsilon+O(\epsilon^2)\simeq
0.35 - 0.10\,\epsilon.
\end{equation}
The probability of winning in this randomised combination of games
A and B is
\begin{eqnarray}
p_{\text{win,AB}} &=& \pi^{\rm st}_{\rm
0AB}\frac{p_{\text{bad}}+p_A}{2}+\left[1-\pi^{\rm st}_{\rm
0AB}\right] \frac{p_{\text{good}}+p_A}{2} \nonumber \\
&=& \frac{727}{1418}-\frac{486795}{5 02681}\epsilon+O(\epsilon^2)
\end{eqnarray}
which is greater than $1/2$ for a sufficiently small $\epsilon$.

This is the mechanism behind the paradox which we have termed
``reorganisation of trends": although game A consists itself in a
negative trend because it uses a slightly bad coin, it increases
the probability of using the good coin of B, i.e., game A
reinforces the positive trend already present in B enough to make
the combination win.

Periodic sequences can also be studied as Markov chains and their
probability of winning in a whole period can be easily computed
using different combinations of the matrices $\Pi_A$ and $\Pi_B$.
Finally, the slopes of the curves in Fig.~\ref{paradoja} can be
calculated as $\langle X(t+1)\rangle-\langle X(t)\rangle
=2p_{\text{win}}-1$.

\subsection{Capital-independent games}
\label{sec:capitalindep}

 The modulo rule in game B is quite
natural in the original representation of the games as a Brownian
ratchet. However, the rule may not suit some applications of the
paradox to biology, biophysics, population genetics, evolution,
and economics. Thus, it would be desirable to devise new
paradoxical games based on rules independent of the capital.
Parrondo, Harmer and Abbott introduced such a game in Ref.
\cite{npg}, inspired by the reorganisation of trends explained in
the last section.

In the new version, game A remains the same as before, but a game
B$'$, which depends on the history of wins and losses of the
player, is introduced. Game B$'$ is played with four coins $B_1'$,
$B_2'$, $B_3'$, $B_4'$ following history-based rules explained in
table \ref{tablahistoria}.
\begin{table}[h]
\begin{center}
\begin{tabular}{ccccc}
Before last & Last & Coin & Prob. of win & Prob. of loss\\
$t-2$&  $t-1$ & at $t$ & at $t$ & at $t$\\
\hline
Loss & Loss & $B_1'$& $p_1$ &$1-p_1$\\
Loss & Win & $B_2'$& $p_2$ &$1-p_2$\\
Win & Loss & $B_3'$& $p_3$ &$1-p_3$\\
Win & Win & $B_4'$& $p_4$ &$1-p_4$\\
\end{tabular}
\end{center}
\caption{History-based rules for game B'}\label{tablahistoria}
\end{table}

\begin{figure}
\begin{center}
\includegraphics[width=0.65\textwidth]{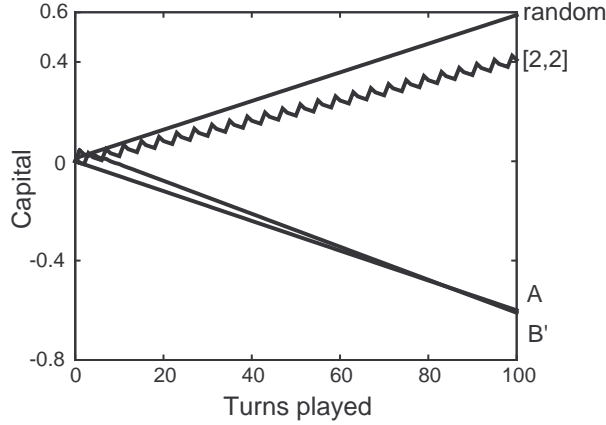}
\caption{Average capital as a function of the number of turns in
the capital independent games. We plot the result for game A and
B$'$, as well as for the random combination and the periodic
sequence AAB$'$B$'$... In all the cases,
$\epsilon=0.003$.}\label{npg}
\end{center}
\end{figure}

The paradox reappears, for instance, when setting
$p_1=9/10-\epsilon$, $p_2=p_3=1/4-\epsilon$, and
$p_4=7/10-\epsilon$. With these numbers and for $\epsilon$ small
and positive, B$'$ is a losing game, while either a random or a
periodic alternation of A and B$'$ produces a winning result.
Fig.~\ref{npg} shows a theoretical computation of the average
capital for these history-dependent paradoxical games.

The paradox is reproduced because there are bad coins in game B$'$
which are played more often than in a completely random game,
i.e., a quarter of the time. For the above choices of $p_i$,
$i=1,2,3,4$, the bad coins are $B_2'$ and $B_3'$. The other two
coins, $B'_1$ and $B'_4$, are good coins.

Due to the fact that game B$'$ rules depend on the history of wins
and losses, the capital $X(t)$ is no longer a Markovian process.
However, the random vector
\begin{equation}
Y(t)=\left(
\begin{array}{c}
X(t)-X(t-1)\\
X(t-1)-X(t-2)
\end{array}
\right)
\end{equation}
can take on four different values and is indeed a Markov chain.
The transition probabilities are again easily obtained from the
rules of game B$'$ and an analytical solution can be obtained
following a similar argument as in section \ref{closerlook} (see
however Ref.  \cite{npg} for details).

We see that the mechanism that we have called reorganisation of
trends can be used to extend the paradox to other gambling games.
It is also noteworthy that the price we must pay to eliminate the
dependence on the capital in the original paradox is to consider
history-dependent rules, i.e., games where the capital is no
longer Markovian.

\section{Collective games}
\label{sec:colect}

In this section we analyse three different versions of paradoxical
games played by a large number of individuals. The three share the
feature that it can sometimes be better for the players to
sacrifice short term benefits for higher returns in the future.

\subsection{Capital redistribution brings wealth.}
\label{sec:raul}

Reorganisation of trends tells us that the essential role of game
A in the paradox is to randomise the capital and make its
distribution more uniform. Toral has found that a redistribution
of the capital in an ensemble of players has the same effect
\cite{capital}.

\begin{figure}
\begin{center}
\includegraphics[width=0.65\textwidth]{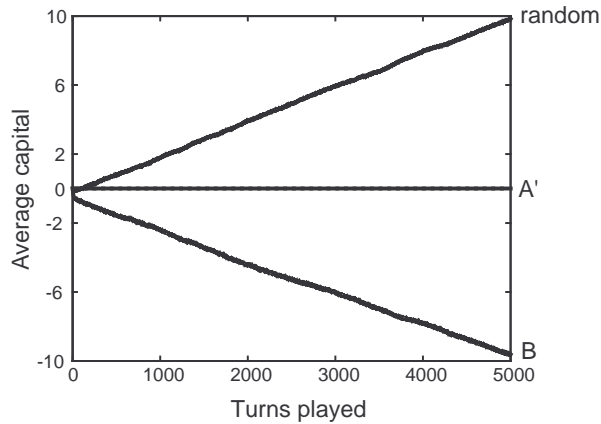}
\end{center}
\caption{Average capital per player as a function of the number of
turns in game B, game A$'$ (redistribution of capital), and the
random combination. The data have been obtained for
$\epsilon=0.01$, simulating an ensemble of 10 players and
averaging over 500 realizations.} \label{simulacion}
\end{figure}

 In the new paradoxical games introduced by Toral in
\cite{capital}, there are $N$ players and one of them is randomly
selected in each turn. They can play two games. The first one,
game A', consists of giving a unit of his capital to another
randomly chosen player in the ensemble, that is, game A$'$ is
nothing but a redistribution of the total capital. The second one,
game B, is the same as in the original paradox. Under game A$'$
the capital does not change, where game B is, as before, a losing
game. The striking result is that the random combination of the
two games is winning, i.e., the redistribution of capital
performed in the turns where A$'$ is played turns the losing game
B into a winning one, actually increasing the total capital
available. Thus, the redistribution of capital turns out to be
beneficial for everybody. This effect is shown in
Fig.~\ref{simulacion} where the average total capital in a
simulation with  10 players and 500 realizations is depicted for
games B and A$'$, and for their random combination. It is
remarkable that the effect is still present when the capital is
required to flow from richer to poorer players (see \cite{capital}
for details).

The explanation to this phenomenon follows the same lines as in
the original paradox.

\subsection{Dangerous choices I: The voting paradox}
\label{ec:voting}

Up to now, we have considered sequences of games that are
``imposed" to the player or players. Either they play game A, game
B, or a periodic or random sequence, but we never allow the
players to choose the game to be played in each turn. In the case
of a single player this deference is quite generous, since her
capital would increase in average under the following trivial
choice: she selects game B if her capital is not a multiple of
three and game A otherwise. This is undoubtedly the best strategy,
because the best coin is always used in every turn. Moreover, it
is not difficult to see that this choice strategy performs much
better than any other random or periodic combination of games.

However, things change when we consider an ensemble of players.
How can the ensemble choose the game to be played in each turn?
There are some possibilities, such as letting them vote for the
preferred game or trying to maximise the winning probability in
each turn. Which is then the best choice strategy? We will see
that the paradoxical games also yield some surprises in this
context: the choice that prefers the majority of the ensemble
turns to be worse than a random or periodic combination of games.
Even if we select the game maximising the profit in every turn, we
can end with systematic loses, as shown in the next section.

Consider a set of $N$ players who play game A or B against a
casino. In each turn, {\em all} of them play the {\em same} game.
Therefore, they have to make a collective decision, choosing
between game A or B in each turn. We will firstly  use a {\em
majority rule}  to select the game, that is, the game which
receives more votes is played by all the players simultaneously.
The vote of each player will be determined by her capital,
following the strategy that we have explained above for a single
player. Players with capital multiple of three will vote for game
A, whereas the rest will vote for B.

\begin{figure}
\begin{center}
\includegraphics[width=0.65\textwidth]{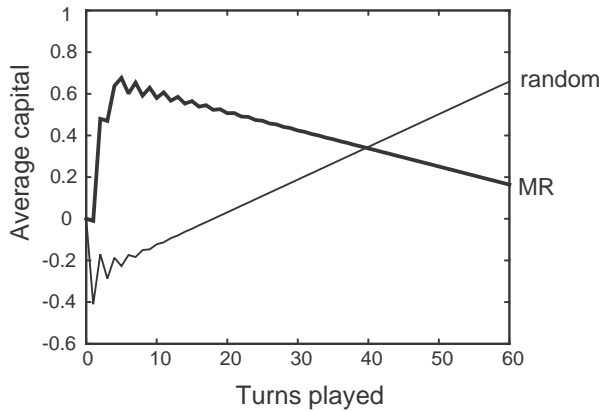}
\caption{Average capital per player in the collective games as a
function of the number of turns, when the game is selected at
random and following the preference of the majority of the players
(MR). Notice that, in the stationary regime, the majority rule
(MR) yields systematic loses whereas the random choice wins in
average. These are analytical results with $\epsilon=0.005$ and an
infinite number of players.} \label{capitalmayoria}
\end{center}
\end{figure}

This strategy, which is optimal for a single player, turns to be
losing if the number of players is large enough \cite{physicaA}.
This is shown in figure \ref{capitalmayoria}, where we plot the
average capital in an ensemble with an infinite number of players.
On the other hand, if the game is selected at random  the capital
increases in time.

In order to explain this behaviour, we will focus on the evolution
of $\pi_0(t)$,  the fraction of players whose capital is a
multiple of three. The selection of the game by voting can be
rephrased in terms of $\pi_0(t)$. As mentioned above, every player
votes for the game which offers him the higher probability of
winning according to his own state. Then, every player whose
capital is a multiple of three  will vote for game A in order to
avoid the bad coin in B. That accounts for a fraction $\pi_0(t)$
of the votes. The remaining fraction $1-\pi_0(t)$ of the players
will vote for game B to play with the good coin. Since the
majority rule establishes that the game which receives more votes
is selected, game A will be played if $\pi_0(t)>1/2$. Conversely,
the whole set of players will play game B when $\pi_0(t)$ is below
$1/2$.

On the other hand, as we have seen in section \ref{closerlook},
playing game B makes $\pi_0(t)$ tend to a stationary value given
by Eq.~(\ref{piob}), namely, $\pi_{\rm 0B}^{\rm st}\simeq 0.38 -
0.2\epsilon<1/2$ for $\epsilon>0$, whereas playing game A makes
$\pi_0$ tend to $1/3$. This is still valid for the present model,
since the $N$ players only interact when they make the collective
decision, otherwise they are completely independent.

\begin{figure}
\begin{center}
\includegraphics[width=0.65\textwidth]{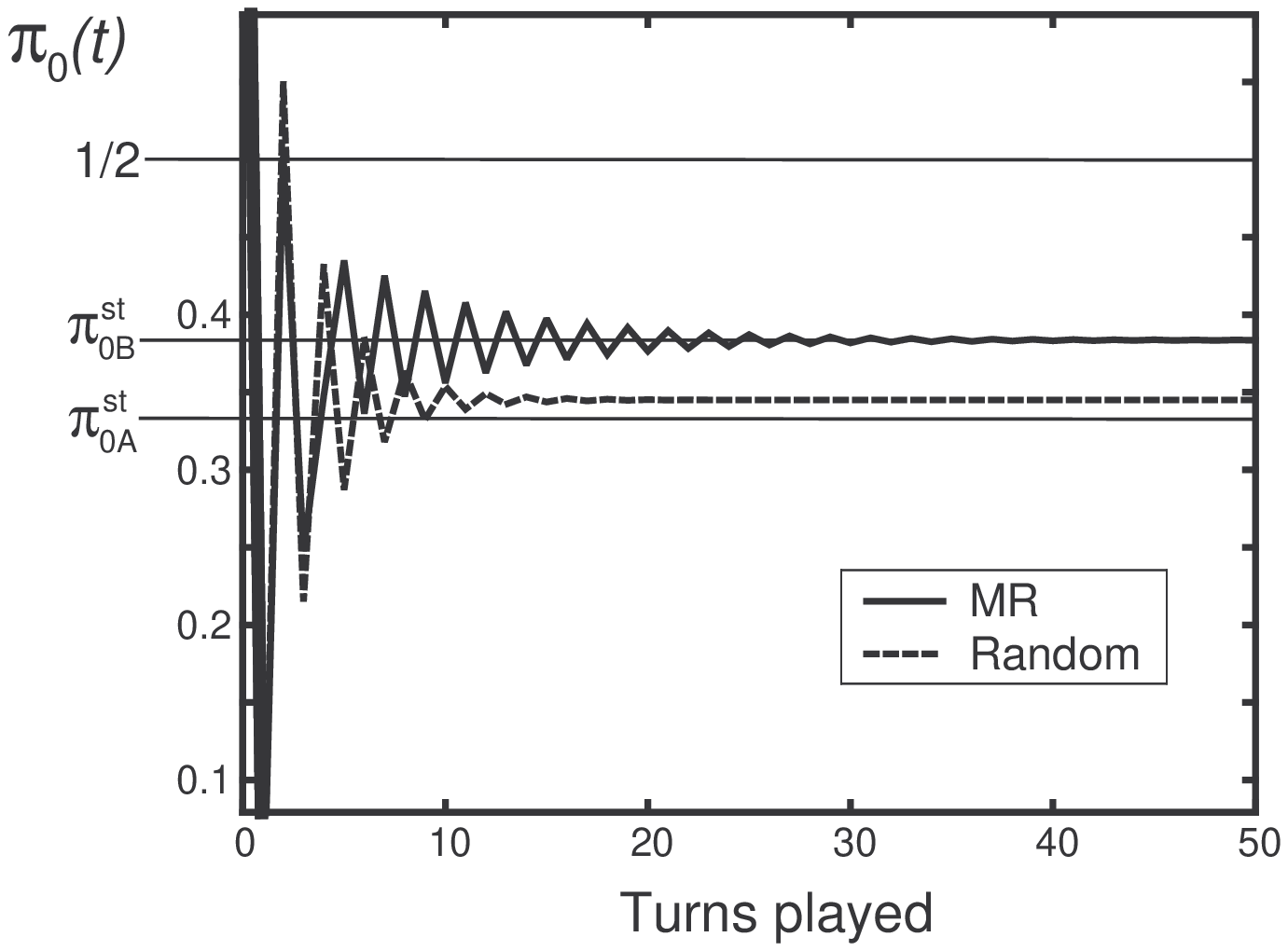}
 \caption{The fraction of players $\pi_0(t)$ with capital multiple of three
 as a function of time
  when the game is chosen at random and following the majority rule (MR).
  In both cases, $\epsilon=0.005$ and $N=\infty$.
  The horizontal lines
indicate the threshold value for the majority rule (1/2), and the
stationary values for games A and B, $\pi_{\rm 0A}^{\rm st}$ and
$\pi_{\rm 0B}^{\rm st}$, respectively. The figure clearly shows
that the random strategy keeps $\pi_0(t)$ small, whereas the
majority rule, selecting B most of the time, drives $\pi_0(t)$ to
a value where both game A and B are losing.} \label{pi0mayoria}
\end{center}
\end{figure}

If $\pi_0(t)>1/2$, then the ensemble of players will select game
A. The fraction $\pi_0(t)$ will decrease until it crosses this
critical value 1/2. At that turn, B is the selected game and it
will remain so as long as $\pi_0$ does not exceed $1/2$. However,
this can never happen, since game B drives $\pi_0$ closer and
closer to $\pi_{\rm 0B}^{\rm st}$ which is below $1/2$. Hence,
after a number of turns, the system gets trapped playing game B
forever with $\pi_0$ asymptotically approaching $\pi_{\rm 0B}^{\rm
st}$. Since $\epsilon$ is positive, game B is a losing game (c.f.~
section \ref{closerlook}) and, therefore, the majority rule yields
systematic losses, as  can be seen in Fig.~\ref{capitalmayoria}.
We have also plotted in Fig.~\ref{pi0mayoria} the fraction
$\pi_0(t)$, to check that, once $\pi_0(t)$ crosses 1/2, game B is
always chosen and $\pi_0(t)$ approaches $\pi_{\rm 0B}^{\rm st}$,
staying far below 1/2.

On the other hand, if, instead of using the majority rule, we
select the game at random or following a periodic sequence, game A
will be chosen even though $\pi_0 < 1/2$. This is a bad choice for
the majority of the players, since playing B would make them toss
the good coin. That is, the random or periodic selection will
contradict from time to time the will of the majority.
Nevertheless, choosing the game at random keeps $\pi_0$ away from
$\pi_{\rm 0B}^{\rm st}$, as shown in Fig.~\ref{pi0mayoria}, i.e.,
in a region where game B is  winning  ($\pi_0<\pi_{\rm 0B}^{\rm
st}$). Therefore, the random choice yields systematic gains, as
shown in Fig.~\ref{capitalmayoria}.

It is worth noting that choosing the game at random is exactly the
same as if every player voted at random. Therefore, the players
get a winning  tendency when they vote at random whereas they lose
their capital when they vote according to their own benefit in
each run.

\subsection{Dangerous choices II: The risks of short-range optimisation}
\label{sec:sr}

Yet another ``losing now to win later'' effect can be observed in
the collective paradoxical games with another choice strategy. As
in the previous example, we consider a large set of players, but
we have to add a small ingredient to achieve the desired effect:
now only a randomly selected fraction $\gamma$ of them play the
game in each turn. Suppose we know the capital of every player so
we can compute which game, A or B, will give the larger average
payoff in the next turn. Again, and even more strikingly,
selecting the ``most favorable game'' results in systematic losses
whereas choosing the game at random or following a periodic
sequence steadily increases the average capital \cite{eplsr}.

\begin{figure}
\begin{center}
\includegraphics[width=0.65\textwidth]{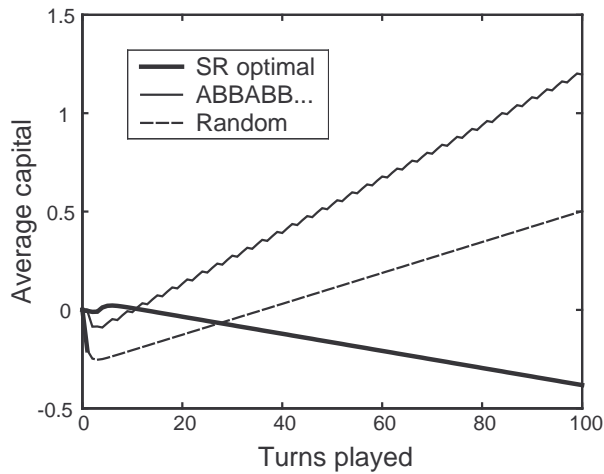}
\caption{Average capital as a function of time for the three
different strategies explained in the text, with $N=\infty$,
$\gamma=0.5$, and $\epsilon=0.005$. The short-range (SR) optimal
strategy is losing in the stationary regime, whereas the two blind
strategies: choosing the game to be played either at random or
following the periodic sequence (ABBABB...), yield  a systematic
gain. } \label{capitalSRepsilon}
\end{center}
\end{figure}

The knowledge of the capital of every player allows us to choose
the game with the highest average payoff in the next turn, since
this optimal game can easily be obtained from  the fraction
$\pi_0(t)$ of players whose capital is a multiple of three. These
individuals will play the bad coin if game B is chosen and the
remaining fraction $1-\pi_0(t)$ will play the good coin. Hence,
the probability of winning for game B reads
\begin{equation}
p_{\text{winB}}=\pi_0p_{\text{bad}}+(1-\pi_0)p_{\text{good} }.
\end{equation} In case game A is selected, the probability to win is
 $p_{\text{winA}}=p_A=1/2-\epsilon$ for all
time $t$. Therefore, to choose the game with the larger payoff
$\langle X(t+1)\rangle-\langle X(t)\rangle =2p_{\text{win}}-1$ in
every turn $t$, we must
\begin{eqnarray}
\text{play A} &\text{if}& p_{\text{winA}}\geq
p_{\text{winB}}(\pi_0)\nonumber\\ \text{play B}&\text{if}&
p_{\text{winA}}< p_{\text{winB}}(\pi_0)
\end{eqnarray}
or equivalently
\begin{eqnarray}
\text{play A} &\text{if}&  \pi_{0}(t)\geq
\pi_{0c}\nonumber\\
\text{play B} &\text{if} & \pi_0(t) < \pi_{0c}
\label{prescription}
\end{eqnarray}
with
$\pi_{0c}\equiv(p_A-p_{\text{good}})/(p_A-p_{\text{bad}})=5/13$.
We will call this way of selecting the game {\em the short-range
optimal strategy}. We will also consider that the game is selected
following either a random or periodic sequence. These are both
{\em blind} strategies, since they do not make any use of the
information about the state of the system. However, and
surprisingly enough, they turn out to be much better than the
short-range optimal strategy, as shown in
Fig.~\ref{capitalSRepsilon}.

\begin{figure}
\begin{center}
\includegraphics[width=0.65\textwidth]{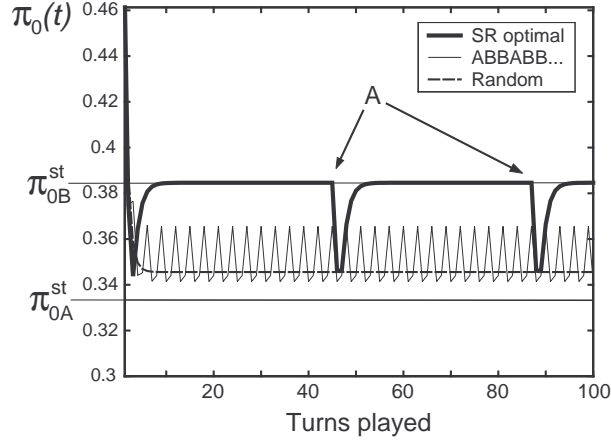}
\caption{The fraction $\pi_0(t)$ of players with capital multiple
of three as a function of the number of turns,
 for $\epsilon=0$, $N=\infty$, and $\gamma=0.5$. The
horizontal lines show the stationary values for game A and game B
(which coincides with the critical fraction $\pi_{0c}$ for the
short-range optimal strategy). As we have in figure
\ref{pi0mayoria} with the majority rule, the short-range optimal
strategy drives $\pi_0(t)$ towards higher values than the other
two strategies.} \label{pi0SR}
\end{center} \end{figure}

\begin{figure}
\begin{center}
\includegraphics[width=0.65\textwidth]{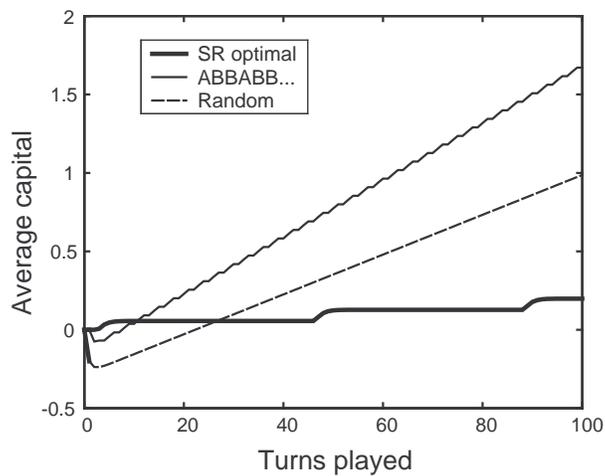}
\caption{Average capital as a function of time for the three
different strategies explained in the text and for $\epsilon=0$,
$N=\infty$ and $\gamma=0.5$. In this case the short-range optimal
strategy is still winning, due to the small jumps coinciding with
the selection of game A. However, most of the turns game B is
played with a value of $\pi_0(t)$ very close to $\pi_{0B}^{\rm
st}$.} \label{capitalSR}
\end{center}
\end{figure}

Notice that \eqref{prescription} is similar to the way the game is
selected by the majority rule considered in the previous section,
but replacing $1/2$ by the new critical value $\pi_{0c}=5/13$.
Therefore, the explanation of this model goes quite along the same
lines as for the voting paradox, although with some differences.
Unlike $1/2$, $\pi_{0c}$ equals the stationary value of $\pi_0(t)$
for game B when $\epsilon=0$. As in the voting paradox, game A
drives $\pi_0$ below $\pi_{0c}$ because game A makes $\pi_0$ tend
to $1/3$. If $\pi_0(t)<\pi_{0c}$, then game B is played, but
$\pi_0(t+1)$ will be still below $\pi_{0c}$ only for $\gamma$
sufficiently small. For example, if $\gamma=1/2$ and $\epsilon=0$,
game B is chosen forty times in a row before switching back to
game A, making $\pi_0$ become approximately equal to $\pi_{\rm
0B}^{\rm st}$ at almost every turn. This behaviour is shown in
Fig.~\ref{pi0SR}. As long as $\pi_0$ is close to $\pi_{\rm
0B}^{\rm st}$, the average capital remains approximately constant,
as shown in Fig.~\ref{capitalSR}.

In contrast, the periodic and random strategies choose game A with
$\pi_0<\pi_{0c}$. Although this does not produce earnings in that
turn, it keeps $\pi_0$ away from $\pi_{\rm 0B}^{\rm st}$. When
game B is chosen again, it has a large expected payoff since
$\pi_0$ is not close to $\pi_{\rm 0B}^{\rm st}$. By keeping
$\pi_0$ not too close to $\pi_{\rm 0B}^{\rm st}$, the blind
strategies perform better than the short-range optimal
prescription, as can be seen in Fig.~\ref{capitalSR}.

The introduction of $\epsilon>0$ has two effects. First of all, it
makes $\pi_{\rm 0B}^{\rm st}$ decrease by a small amount, as
indicated in Eq.~(\ref{piob}). This makes it even more difficult
for the short-range optimal strategy to choose game A, and after a
few runs game B is always selected. Since game B is now a losing
game, the short-range optimal strategy is also losing whereas
periodic and random strategies keep their winning tendency, as can
be seen in Fig.~\ref{capitalSRepsilon}.

To summarise, the short-range optimal strategy chooses B most of
the time, since it is the game which gives the highest returns in
each turn. However, this choice drives $\pi_0(t)$ to a region in
which B is no longer a winning game. On the other hand, the random
strategy from time to time sacrifices the short term returns by
selecting game A, but this choice keeps the system in a
``productive region". We could say that the short-range optimal
strategy is ``killing the goose that laid the golden eggs", an
effect that is also present in simple deterministic systems
\cite{eplsr}.

\section{Conclusions}
\label{sec:conclusions}

We have presented the original Parrondo's paradox and several
examples showing how the basic mechanisms underlying the paradox
can yield other counter-intuitive phenomena. We finish by
reviewing these mechanisms as well as the literature related with
the paradox.

The first mechanism, the ratchet effect, occurs when fluctuations
can help to surmount a potential barrier or a ``losing streak''.
These fluctuation can either come from another losing game, such
as in the original paradox, from a redistribution of the capital,
such as in Toral's collective games \cite{capital}, or from a
purely diffusive motion, such as in the flashing ratchet.

A second mechanism is the reorganisation of trends, which occurs
when game A reinforces a positive trend already present in game B.
The same mechanism can be observed in the games with capital
independent rules and it helps to understand the counter-intuitive
behaviour of the collective games presented in section
\ref{ec:voting} and \ref{sec:sr}, where random choices perform
better than the choice preferred by the majority or the one
optimizing short term returns. These models also prompt the
question of how information can be used to design a strategy. It
is a relevant question for control theory and also for statistical
mechanics, since the paradox is a purely collective effect that
goes away for a single player, i.e., the choices following the
short-range optimal strategy and the majority rule perform much
better than the random or periodic ones.

There is a third mechanism which we have not addressed along the
paper, but immediately arises if we consider the games as
dynamical systems: the outcome of an alternation of dynamics can
always be interpreted as a stabilization of transient states. This
interpretation has allowed some authors to extend the basic
message of the paradox to pattern formation in spatially extended
systems \cite{prl,pre,fnl,turing}. In these papers, a new
mechanism of pattern formation based on the alternation of
dynamics is introduced. They show how the global alternation of
two dynamics, each of which leads to a homogeneous steady state,
can produce stationary or oscillatory patterns upon alternation.

Another interesting application of the stabilization of transient
states is presented in Ref. \cite{hanta}. Two dynamics for the
population of a virus are introduced with the following property:
in each dynamics, the population vanishes, whereas the alternation
of the two dynamics, whose origin could be the seasonal variation,
induces an outbreak of the virus.

Similar effects can be seen in quantum systems. Lee {\em et al.}
have devised a toy model in which the alternation of two
decoherence dynamics can significantly decrease the decoherence
rate of each separate dynamics \cite{lee}. Also in the quantum
domain, the paradox has received some attention: there have been
some proposals of a quantum version of the games
\cite{abbotq,meyer2} closely related with the recent theory of
quantum games \cite{eisert}, and the paradox has been reproduced
in the contexts of quantum lattice gases \cite{meyer1} and quantum
algorithms \cite{lee2}.

To finish this partial account of the existing literature on the
paradox, we mention the work by Arena {\em et al} \cite{chaot},
who analyse the performance of the games using chaotic instead of
random sequences of choices; that of Chang and Tsong \cite{trunc},
who study the hidden coupling between the two games in the paradox
and present several extensions even for deterministic dynamics;
and the paper by Kocarev and Tasev \cite{lyap}, relating the
paradox with Lyapunov exponents and stochastic synchronisation.

In summary, Parrondo's paradox has drawn the attention of many
researchers to non-trivial phenomena associated with switching
between two dynamics.  We have tried to reveal in this paper some
of the basic mechanisms that can yield an unexpected behaviour
when switching between two dynamics, and how these mechanisms work
in several versions of the paradox. As mentioned in the
introduction, we believe that the paradox and its extensions are
contributing to a deeper understanding of stochastic dynamical
systems. In the case of statistical mechanics, switching is in
fact a source of non-equilibrium which is ubiquitous in nature,
due to day-night or seasonal variations \cite{hanta}.
Nevertheless, it has not been studied in depth until the recent
introduction of ratchets and paradoxical games. As the paradox
suggests, we will probably see in the future new models and
applications confirming that noise and switching, even between
equilibrium dynamics, can be a powerful combination to create
order and complexity.

\section*{Acknowledgements}

 We thank valuable comments on the original manuscript made by Katja Lindenberg, Javier
 Buceta, Martin Plenio, and H. Leonardo Mart\'{\i}nez. This work was
supported by a grant from the {\em New Del Amo Program} (U.C.M.),
and by MCYT-Spain Grant BFM 2001-0291.

\section*{Biographies}

Juan M.R. Parrondo  studied in the Universidad Complutense de
Madrid (Spain) where he obtained both his Major (1987) and his PhD
(1992) in Physics. He worked as postdoc in the University of
California, San Diego (USA) and in the Limburgs Universitair
Centrum (Belgium). Since 1996 he is professor in the Universidad
Complutense de Madrid. His main research interests are the the
study of fluctuations in non equilibrium statistical physics and
the foundations of statistical mechanics.

Luis Din\'{\i}s obtained his Major in Physics (2000) and his
Master in Complex Systems (2002) in the Universidad Complutense de
Madrid. His main research interests lie in the area of statistical
physics and biophysics.

\end{document}